\documentclass[preprint,aps,showpacs,nofootinbib,preprintnumbers,amsmath,amssymb]{revtex4-1}
\usepackage{}
\usepackage{epsfig}
\usepackage{subfigure}
\usepackage{dcolumn}
\usepackage{bm}
\usepackage[usenames ,dvipsnames]{xcolor}
\usepackage{slashed}
\usepackage{graphicx,color}

\makeatletter

\@ifundefined{textcolor}{}
{%
 \definecolor{BLACK}{gray}{0}
 \definecolor{WHITE}{gray}{1}
 \definecolor{RED}{rgb}{1,0,0}
 \definecolor{GREEN}{rgb}{0,1,0}
 \definecolor{BLUE}{rgb}{0,0,1}
 \definecolor{CYAN}{cmyk}{1,0,0,0}
 \definecolor{MAGENTA}{cmyk}{0,1,0,0}
 \definecolor{YELLOW}{cmyk}{0,0,1,0}
 }

\begin{document}

\title{Matter Power Spectra in Viable $f(R)$ Gravity Models with Massive Neutrinos }

\author{Chao-Qiang Geng$^{1,2,3}$\footnote{geng@phys.nthu.edu.tw}, Chung-Chi Lee$^3$\footnote{g9522545@oz.nthu.edu.tw},
Jia-Liang Shen$^2$\footnote{dddirac@gmail.com}}
\affiliation{$^1$Chongqing University of Posts \& Telecommunications, Chongqing, 400065, China\\
$^2$Department of Physics, National Tsing Hua University, Hsinchu, Taiwan 300\\
$^3$National Center for Theoretical Sciences, Hsinchu, Taiwan 300}

\date{\today}

\begin{abstract}
We investigate the matter power spectra in the power law and exponential types of viable $f(R)$ theories along with massive neutrinos.
The enhancement of the matter power spectrum is found to be a generic feature in these models.
In particular, we show that in the former type, such as the Starobinsky model, the spectrum is magnified
much larger than the latter one, such as the exponential model. A greater scale of  the total neutrino mass, $\Sigma m_{\nu}$,
is allowed in the viable $f(R)$ models than that in the $\Lambda$CDM one.
We obtain the constraints on the neutrino masses by using the CosmoMC package with the modified MGCAMB.
Explicitly, we get  $\Sigma m_{\nu} < 0.451 \ (0.214)\ \mathrm{eV}$ at $95\%$ C.L. in the Starobinsky (exponential) model,
while the corresponding one for the $\Lambda$CDM model is $\Sigma m_{\nu} < 0.200\ \mathrm{eV}$.
Furthermore, by treating the effective number of neutrino species $N_{\mathrm{eff}}$  as a free parameter along with $\Sigma m_{\nu}$,
we  find that $N_{\mathrm{eff}} = 3.78^{+0.64}_{-0.84}~(3.47^{+0.74}_{-0.60})$ and $\Sigma m_{\nu} = 0.533^{+0.254}_{-0.411}$
 ($< 0.386) \ \mathrm{eV}$ at $95\%$ C.L.  in the Starobinsky (exponential) model.
\end{abstract}



\maketitle

\section{Introduction}

The observations of the Type-Ia supernovae~\cite{Riess:1998cb, Perlmutter:1998np} in the last decade of the 20th century
indicated that our universe is undergoing an accelerating expansion. Since then, the phenomenon has been further verified
by several succeeding experiments~\cite{Spergel:2003cb, Ade:2013zuv, Tegmark:2003ud}. To explain this interesting phenomenon,
people have tried various methods. One of which is to introduce a homogeneous and isotropic energy density with negative pressure
into the theory of General Relativity (GR), so-called  ``Dark Energy.''\cite{DE}  The other way is to modify Einstein's gravity theory
by extending the Ricci scalar $R$ in the Einstein-Hilbert action to an arbitrary function $f(R)$~\cite{frxx, arXiv:1002.4928}.
Several viable $f(R)$ models have been proposed to satisfy the constraints
from theoretical considerations as well as cosmological observations~\cite{arXiv:1002.4928}.

On the other hand, the oscillations between the three  flavors of neutrinos in the standard model of particle physics have
been detected~\cite{Aguilar:2001ty, AguilarArevalo:2010wv}, suggesting that either two or  three of the active neutrinos have tiny masses.
Clearly, from the cosmological point of view, it is necessary to consider the effect of massive neutrinos on the evolution of
our universe~\cite{Lesgourgues:2006nd,Motohashi:2010sj,Motohashi:2012wc, He:2013qha, Dossett:2014oia, Zhou:2014fva}.
For example, massive neutrinos will suppress the matter power spectrum in the small scale~\cite{Lesgourgues:2006nd,Motohashi:2010sj}.
In other words, cosmology offers  strong constraints on the mass scales of neutrinos.
However, such cosmological constraints are highly model dependent. It is know
that  the simplest model in cosmology, the $\Lambda$CDM model, permits only a small range for the sum of the active neutrino masses.
For example, the constraint from Planck~\cite{Ade:2013zuv} allows $\Sigma m_{\nu} < 0.23 eV$ at $95\%$ confidence level. 
 In the viable $f(R)$ models,
 their matter power spectra are normally larger than that of the $\Lambda$CDM~\cite{Motohashi:2010sj,Motohashi:2012wc}.
 This enhancement then can be used to compensate
 for the suppression due to massive neutrinos
so that
 the neutrino mass constraint is relaxed to a broader window.
 Recently,
$\Sigma m_{\nu} < 0.5 $ eV has been extracted in
Refs.~\cite{He:2013qha, Dossett:2014oia, Zhou:2014fva}
for the chameleon type of  $f(R)$ gravity.
In this paper, we will examine two typical viable $f(R)$ models with their exact forms.

In general, the viable $f(R)$ models can be categorized into power-law and exponential types.
In this paper, we focus on the Starobinsky and  exponential gravity models, which belong to these two types, respectively.
Without loss of generality, we consider these $f(R)$ models with one massive neutrino along with the other two being massless.
Using the modified Code for Anisotropies in the Microwave Background (MGCAMB)~\cite{Lewis:1999bs, Hojjati:2011ix}
and the CosmoMC package~\cite{Lewis:2002ah}, we study the constraints on the neutrino masses from
the latest cosmological observational data, including those of
 the cosmic microwave background  (CMB)  from Planck~\cite{Ade:2013zuv}
and WMAP~\cite{Hinshaw:2012aka}, baryon acoustic oscillation (BAO)  from
Baryon Oscillation Spectroscopic Survey (BOSS)~\cite{Anderson:2012sa},
Type-Ia supernova (SNIa) from Supernova Legacy Survey (SNLS)~\cite{Astier:2005qq}, and
matter power spectrum  from Sloan Digital Sky Survey (SDSS)~\cite{AdelmanMcCarthy:2005se}
and WiggleZ Dark Energy Survey~\cite{Blake:2011wn}.
The constraint on the effective number of neutrino species, $N_{\mathrm{eff}}$,
is also acquired in order to examine the non-standard properties of neutrinos.
Since MGCAMB uses the parametrized framework to include $f(R)$ gravity into CAMB,
we only consider the linear perturbation and assume that the background evolution is the same as the $\Lambda$CDM model.

This paper is organized as follows: In Sec.~2, we first give a brief review on the $f(R)$ modification in the linear perturbation theory
and then show the matter power spectra $P(k)$ in the two types of  the viable $f(R)$ models. The effect of massive neutrinos is examined.
In Sec.~3, we show the results of the constraints on massive neutrinos using the CosmoMC package.
Finally, we present our conclusions in Sec.~4.

\section{Matter Power Spectrum in $f(R)$ Gravity}

\subsection{$f(R)$ Theory}
The action of $f(R)$ gravity is given by
\begin{eqnarray}
S = \frac{1}{\kappa^2}  \int \sqrt{-g} \ f(R)  d^4x  + \mathcal{L}_M \,,
\label{eq:action-1}
\end{eqnarray}
where $\kappa^2 \equiv 8 \pi G$, $g$ is the determinant of metric tensor,
$f(R)$ is an arbitrary function of Ricci scalar and $\mathcal{L}_M$ denotes
the matter Lagrangian density. By varying the action~(\ref{eq:action-1}) with
respect to the metric $g_{\mu \nu}$, we obtain
\begin{eqnarray}
f_R R_{\mu\nu} - \frac{1}{2}f(R)g_{\mu\nu} + \left( g_{\mu\nu}\Box - \nabla_{\mu}\nabla_{\nu} \right) f_R = \kappa^{2}  T_{\mu\nu} \,,
\label{eq:eom_fR}
\end{eqnarray}
where the subscript ``R'' denotes the derivative of R, $i.e.$ $f_R \equiv \partial f/\partial R$,
$ \Box = g^{\mu \nu} \nabla_{\mu} \nabla_{\nu}$ is the d'Alembertian operator,
 and $T_{\mu \nu}$ is the energy-momentum tensor, defined by
\begin{equation}
T_{\mu\nu} \equiv \frac{-2}{\sqrt{-g}}\frac{\delta S_{m}}{\delta g^{\mu\nu}}\,.
\label{eq:EMtensor}
\end{equation}

To deal with the dark energy problem, we must check whether $f(R)$ gravity satisfies the following viable
conditions: (i) $f_R > 0$ for $R > R_0 \equiv R (z=0)$, which keeps the positivity of the modified
Newton constant and avoids the attractive gravitational force;
(ii) $f_{RR} > 0$ for $R > R_0$, which guarantees that the mass of the scalaron is real defined;
(iii) $f(R) \rightarrow R - 2 \Lambda$ in the high redshift region ($R \gg R_0$), which reproduces
the $\Lambda $CDM behavior in the early universe;
(iv) a late-time stable solution exists, which eliminates the appearance of singularity in the future; and
(v) it should pass the local gravity tests, including those from the equivalence principle and solar system.
Several viable $f(R)$ models  with these conditions have been proposed, such as  Hu-Sawicki~\cite{arXiv:0705.1158},
Starobinsky~\cite{arXiv:0706.2041}, Tsujikawa~\cite{arXiv:0709.1391},
exponential~\cite{astro-ph/0511218, arXiv:0712.4017, arXiv:0905.2962, arXiv:1005.4574},
and Appleby-Battye~\cite{arXiv:0705.3199, arXiv:0909.1737}  gravity models.
These models can be grouped into power law and exponential types, denoted as
Type-I and II, respectively~\cite{Lee:2012dk}.
In the following discussion, in order to investigate the behaviors of these two types of models,
we  concentrate on the Starobinsky and  exponential models,
given by
\begin{eqnarray}
&&f(R) = R - \lambda R_c \left[ 1- \left( 1+ \frac{R^2}{R_c^2} \right)^{-n} \right] \,,~~\mathrm{(Type-I)}
\label{eq:starmodel}
\\
&&f(R) = R- \beta R_c \left( 1- e^{-R/R_c} \right) \,,~~\mathrm{(Type-II)}
\label{eq:expmodel}
\end{eqnarray}
respectively, where $R_c$ represents the characteristic curvature.

\subsection{Matter Power Spectrum}

We review the perturbation equations of $f(R)$ gravity in the Newtonian gauge and study
the effect of $f(R)$ gravity on the matter power spectrum with the parametrization used
in MGCAMB~\cite{Lewis:1999bs, Hojjati:2011ix}.

The perturbed FLRW metric in the Newtonian gauge is given by
\begin{eqnarray}
&&ds^{2} = a^{2}(\eta) \left[ - ( 1 + 2\Psi )d \eta^2 + ( 1 - 2\Phi )\delta_{ij} dx^idx^j \right] \,,
\label{eq:NewtonGauge}
\end{eqnarray}
where $a(\eta)$ is the scale factor, $\eta$ is the conformal time, and $\Psi$ and $\Phi$ are the scalar perturbations.

It is worth noting that the viable $f(R)$ gravity theory is indistinguishable from the $\Lambda $CDM model at the high-redshift stage.
Although the specific distinction between $f(R)$ and $\Lambda $CDM depends on the forms of $f(R)$, it always happened later
than $z = 10$, at which the non-relativistic matter  dominated our universe. Therefore, we only consider the perturbation equations in the matter dominated region. The perturbed energy-momentum tensor is
\begin{eqnarray}
&& T^0_0 = - (\rho_m + \delta \rho_m) \,,
\label{eq:E-M-per00}
\\
&& T^0_i = - (\rho_m + P_M) v_{m,i} \,,
\label{eq:E-M-per0i}
\end{eqnarray}
where $v_m$ is the velocity field. By following the similar procedure in Ref.~\cite{Tsujikawa:2007gd}
with the energy-momentum conservation, $\delta T^{\mu}_{\nu ; \mu} = 0$,
we can derive the perturbation equations in $f(R)$ theory:
\begin{eqnarray}
&&\frac{k^{2}}{a^{2}}\Phi + 3H(H\Psi + \dot{\Phi}) = \frac{1}{2f_R}  \left[ 3H \delta \dot{f}_R
-\left( 3\dot{H} + 3H^{2} - \frac{k^{2}}{a^{2}} \right) \delta f_R \right.
\nonumber \\
&& \qquad \qquad \qquad \qquad \left. - 3H\dot{f}_R \Psi - 3\dot{f}_R (H\Psi + \dot{\Phi})
- \kappa^{2} \delta \rho_m \right] \,,
\label{eq:perurb-1}
\\
&&\delta \ddot{f}_R + 3H \delta \dot{f}_R + \left (\frac{k^{2}}{a^{2}} - \frac{R}{3}\right )\delta
f_R = \frac{\kappa^{2}}{3}\delta \rho_m + \dot{f}_R (3H\Psi + \dot{\Psi} + 3\dot{\Phi})
\nonumber \\
&&\qquad \qquad \qquad \qquad+ (2\ddot{f}_R+3H\dot{f}_R)\Psi - \frac{1}{3}f_R\delta R \,,
\label{eq:perurb-2}
\end{eqnarray}
\begin{eqnarray}
&&\Psi - \Phi = -\frac{\delta f_R}{f_R} \,,
\label{eq:perurb-3}
\\
&&\delta \dot{\rho}_m + 3H\delta \rho_m = \rho_m \left (3\dot{\Phi} - \frac{k^{2}}{a}v_m \right ) \,,
\label{eq:perurb-4}
\\
&&\dot{v}_m + Hv_m = \frac{1}{a}\Psi \,,
\label{eq:perurb-5}
\end{eqnarray}
where the ``dot" denotes the time derivative and $k$ is a comoving wavenumber under the
Fourier transform.
Besides, it is convenient to use the gauge-invariant matter density perturbation $\delta_{m}$, defined by
\begin{eqnarray}
\delta_{m} \equiv \frac{\delta \rho_{m}}{\rho_{m}} + 3 H v \,, \;\;
v \equiv a v_m \,.
\label{eq:gaugeinvdel-1}
\end{eqnarray}

 From Eqs.~(\ref{eq:perurb-1}) - (\ref{eq:gaugeinvdel-1}) with the sub-horizon limit ($k^{2} \gg a^{2}H^{2}$),
the relations between the metric potential and the perturbation of the matter density are given by
\begin{eqnarray}
&&\frac{k^2}{a^2} \Psi = - 4 \pi G_{eff}(k,a) \, \rho_m \delta_m \,,
\label{eq:poisson}
\\
&&\frac{\Phi}{\Psi} = \gamma(k,a) \,,
\label{time-spatial}
\end{eqnarray}
where
\begin{eqnarray}
&&G_{eff} = \frac{G}{f_R} \frac{1+4 \frac{k^2}{a^2} \frac{f_{RR}}{f_R}}{1+3 \frac{k^2}{a^2} \frac{f_{RR}}{f_R}} \,,
\label{eq:Geff}
\\
&&\gamma =  \frac{1+2 \frac{k^2}{a^2} \frac{f_{RR}}{f_R}}{1+ 4 \frac{k^2}{a^2} \frac{f_{RR}}{f_R}} \,.
\label{eq:gamma}
\end{eqnarray}
Eqs.~(\ref{eq:poisson}) and (\ref{time-spatial}) are the parametrizations used in MGCAMB to incorporate  $f(R)$ gravity.
In other words, MGCAMB modifies the linear perturbation to include the effect of $f(R)$ that has the $\Lambda$CDM limit.

In MGCAMB,
 the $f(R)$ gravity effect
 is introduced by Bertschinger and Zukin (BZ) in Ref.~\citep{Bertschinger:2008zb}
 through the form
\begin{eqnarray}
\label{eq:BZform1}
\frac{G_{\mathrm{eff}}}{G} = \frac{1}{1-1.4\times 10^{-8} \vert \lambda_1 \vert^2 a^3} \frac{1+\frac{4}{3} \lambda_1^2 k^2 a^4}{1+ \lambda_1^2 k^2 a^4}\qquad
\end{eqnarray}
where $\lambda_1$ is defined by the Compton wavelength in units of the Hubble scale~\citep{Hojjati:2011ix, Bertschinger:2008zb},
\begin{eqnarray}
\label{eq:Compton}
\lambda_1^2 = \frac{B_0}{2H_0^2}\,,
\end{eqnarray}
where
\begin{eqnarray}
\label{eq:Compton1}
B_0 \equiv \frac{f_{RR}}{f_R} \frac{dR}{d\ln a} \left( \frac{d \ln H}{d \ln a} \right)^{-1} \bigg{\rvert}_{a=1}\,.
\end{eqnarray}
However, comparing Eq.~(\ref{eq:Geff}) to (\ref{eq:BZform1}),
the BZ approach describes the $f(R)$ effect as a function of time, indicating $f_{RR} \propto a^6$ in the high redshift regime ($f_R \sim 1$).
This approach
can only be used in the Starobinsky model with $n=1$ in the matter dominated era, but not
in the Starobinsky ($n \neq 1$) and exponential models
 since they deviate from the $\Lambda$CDM model as $ f_{RR} \propto (R_c/R)^{2n}$  and $e^{-R/R_c}$, respectively.
Instead of the BZ approach used in Refs.~\cite{Dossett:2014oia, Hojjati:2011ix},
we further modify MGCAMB and take the exact $f(R)$ form in Eq.~(\ref{eq:Geff}) to
represent the matter power spectrum in $f(R)$ gravity. 
Note that in Refs.~\cite{He:2013qha, Zhou:2014fva} the authors have done the analysis by using the full linear perturbation equations of
$f(R)$ gravity~\cite{He:2012}.
Through the viable condition (iii), the
characteristic curvature $R_c$ is defined by the dark energy density $\rho_{\Lambda}$ and the parameter $\lambda$ ($\beta$),
in the Starobinsky (exponential) model,
given by
\begin{eqnarray}
\label{eq:Rch}
f(R) \overset{R \gg R_0}{=} R - \lambda R_c (\beta R_c) \simeq R - 2 \Lambda \quad \Rightarrow \quad R_c \simeq \frac{2\Lambda}{\lambda (\beta)}\,.
\end{eqnarray}
Within this framework, we conduct the program of the CosmoMC package with MGCAMB to perform our calculations.

\begin{center}
\begin{figure}[htbp]
\includegraphics[width=0.49 \linewidth]{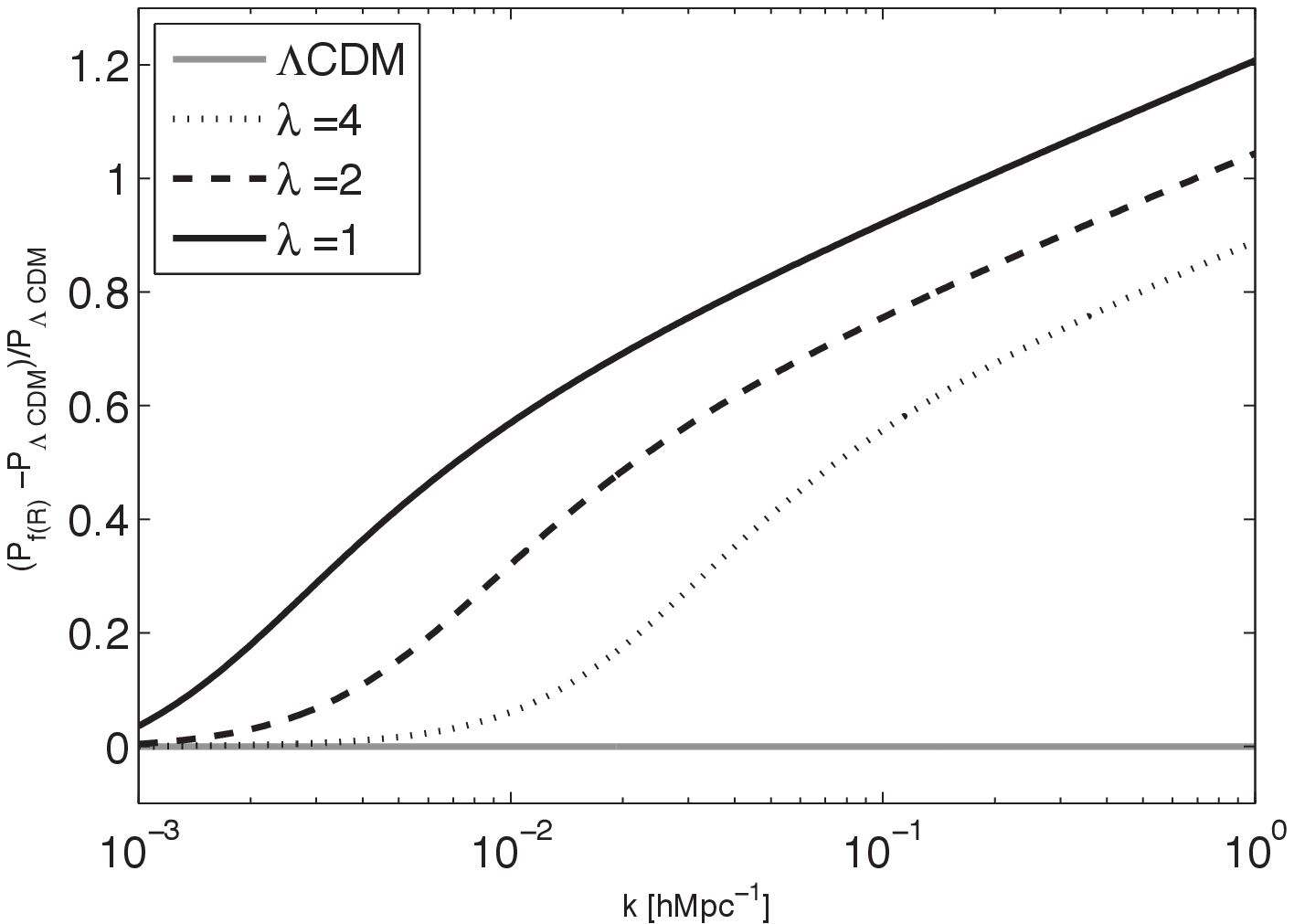}
\includegraphics[width=0.49 \linewidth]{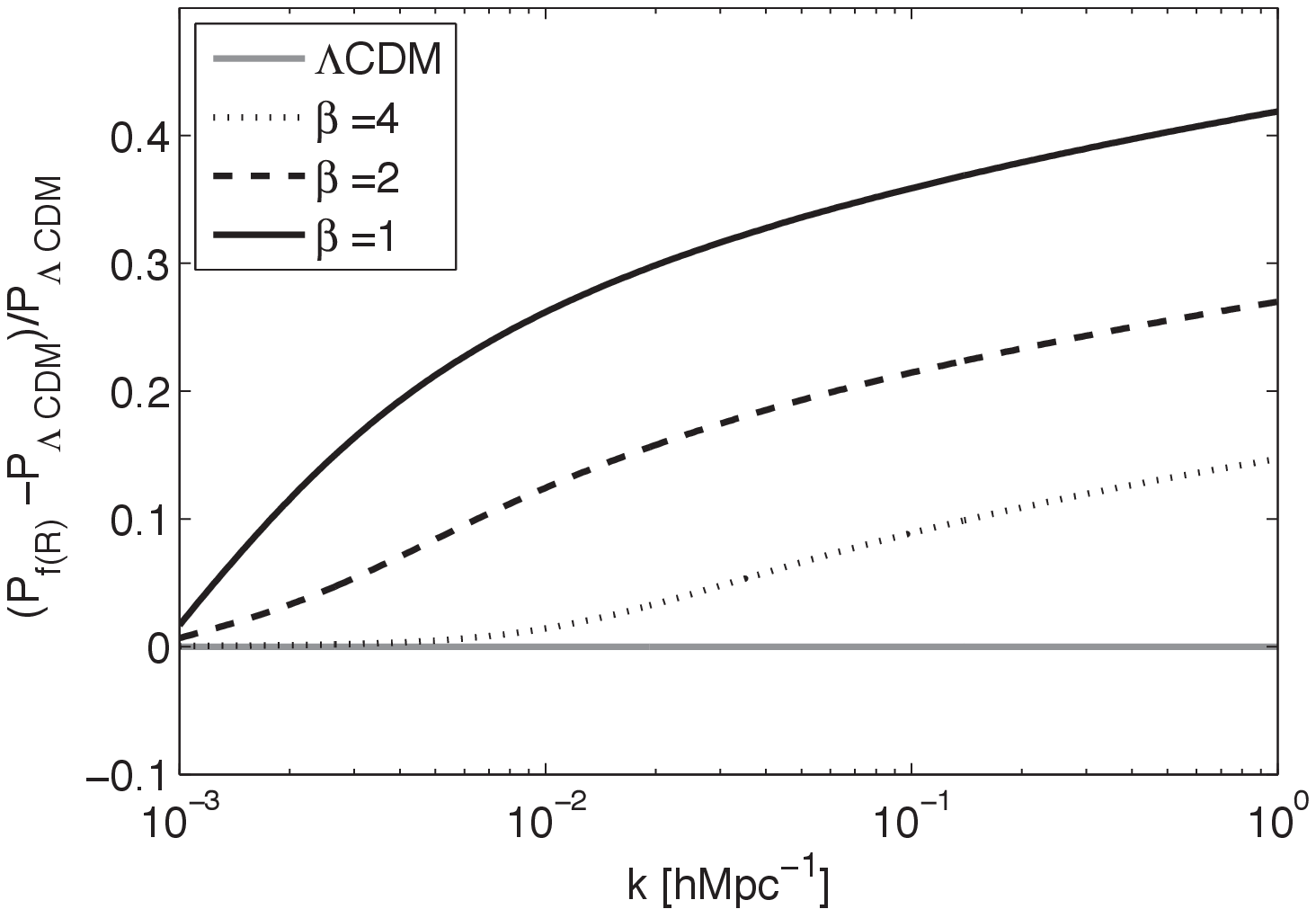}
\caption{
Differences of the matter power spectra  between  $f(R)$ and  $\Lambda$CDM models, where the left and right panels correspond
to the Starobinsky (n=2) and  exponential models, respectively, and the gray solid line represents $P(k)$ in the $\Lambda$CDM model as the scale independent baseline.
}
\label{fg:1}
\end{figure}
\end{center}

Instead of investigating the numerical result from MGCAMB directly, we can figure out the effect of $f(R)$ gravity
on the matter power spectrum, $P(k)\sim \langle \delta_{m}^{2} \rangle$, based on the equation,
\begin{eqnarray}
\ddot{\delta}_m + 2 H \dot{\delta}_m - 4 \pi G_{eff} \rho_m \delta_m = 0 \,,
\label{eq:diffeqofmatter}
\end{eqnarray}
 derived from Eqs.~(\ref{eq:perurb-1}) - (\ref{eq:gaugeinvdel-1}). If we set $f(R) = R - 2 \Lambda$ and $G_{eff} = G$,
 we recover  the $\Lambda$CDM case. According to Eq.~(\ref{eq:diffeqofmatter}) and the definition of $G_{eff}$ in Eq.~(\ref{eq:Geff}),
  the deviation of $\delta_{m}$ from the $\Lambda$CDM model can be estimated by the separation of variables,
\begin{eqnarray}
G_{eff} = G \mu_1(a) \ \mu_2(a,k),
\end{eqnarray}
where $ \mu_1 = f_R^{-1} $ and
$\mu_2 =  \left( 1+4 k^2\, f_{RR}/a^2 \, f_R \right) /
\left( 1+3 k^2\, f_{RR}/a^2 \, f_R \right)$. For the Starobinsky and  exponential models, their first-order derivatives,
\begin{eqnarray}
&&f_R^{\mathrm{(s)}} = 1 - 2 n \lambda \frac{R}{R_c} \left( 1+ \frac{R^2}{R_c^2} \right)^{-(n+1)} \,,
\label{eq:dstar}
\\
&&f_R^{\mathrm{(exp)}} = 1 - \beta e^{-R / R_c} \,,
\label{eq:dexp}
\end{eqnarray}
are both smaller than unity, so that  $\mu_{1,2}$ are greater than unity.
The first two viable conditions, $f_{R} > 0$ and $f_{RR} > 0$, guarantee that the numerator of
$\mu_{2}$ is larger than its denominator. Furthermore, $\mu_{2}$ is enhanced by a large wavenumber $k$,
corresponding to the matter power spectrum in the small scale.
In all, the matter power spectra of the viable $f(R)$ gravity theories are always larger
than that of the $\Lambda$CDM model as explicitly shown in Fig.~\ref{fg:1}, where
the density of dark energy is fixed to be $\Omega_{DE} \sim 73 \% $ and neutrino masses are taken to be zero.

\begin{center}
\begin{figure}[htbp]
\includegraphics[width=0.49 \linewidth]{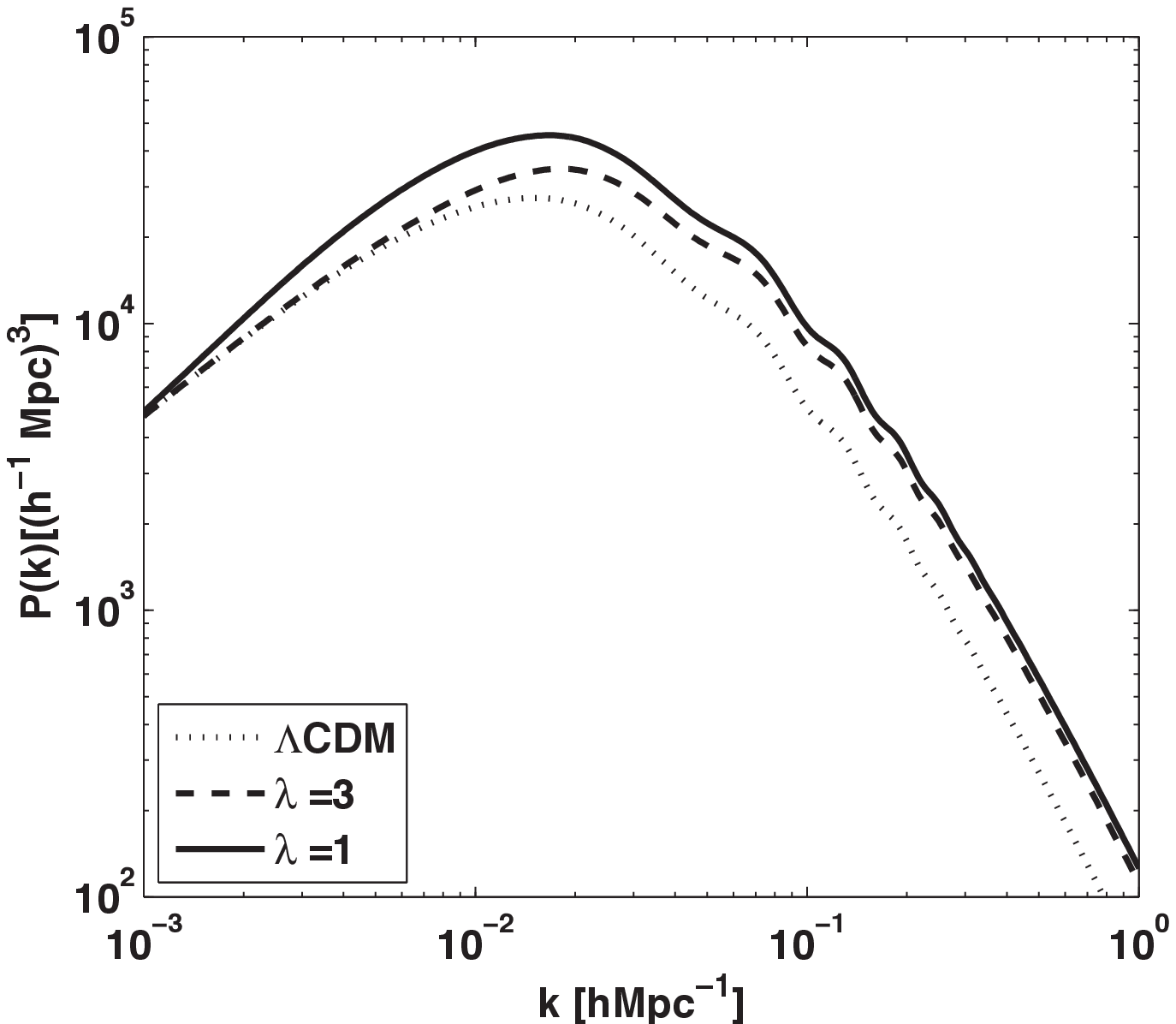}
\includegraphics[width=0.49 \linewidth]{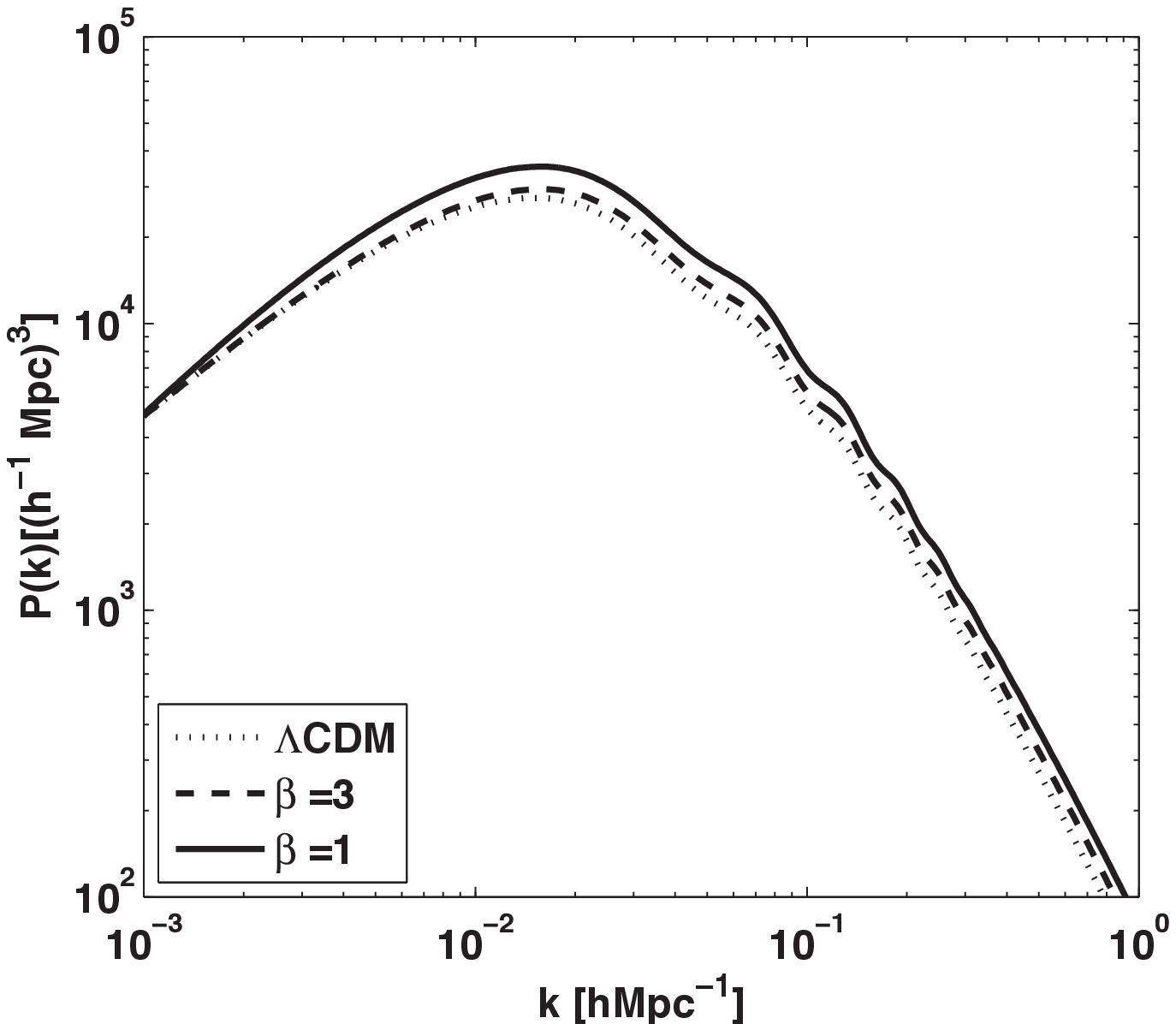}
\caption{
Matter power spectra $P(k)$ of the Starobinsky (left with fixing $n=2$) and exponential (right) models,
where the dotted lines represent $P(k)$ in the $\Lambda $CDM model.
}
\label{fg:2}
\end{figure}
\end{center}

The  deviation between the Starobinsky (exponential) and
 $\Lambda$CDM models is proportional to
$(R_c/R)^{2n}$ ($e^{-R/R_c}$).
If the dark energy density $\rho_{\Lambda}$ is fixed at the present time ($z=0$), the characteristic curvature $R_c$ is inverse proportional to
the model parameter, $R_c \sim 2\Lambda/\lambda(\beta)$, denoting that a smaller model parameter corresponds to
a larger deviation for the  matter power spectrum.
As a result,
we consider the model parameter space  close to the lower bound of the viable condition
($\lambda, \beta \geq 1$ and $n \geq 2$) in the following calculations. 
In Fig.~\ref{fg:2}, we show the matter power spectra in
the Starobinsky $(n=2)$ and exponential models, where the $\Lambda$CDM result is also given.
As indicated above,
a large enhancement of $f(R)$ gravity occurs at the large wavenumber $k$.
There is an interesting phenomenon: the magnification of $P(k)$ in the Starobinsky model is
more greater than that in the exponential one within
the same allowed viable model parameter  (e.g., $\lambda = \beta$ and $R_c^{\mathrm{star}}= R_c^{\mathrm{exp}}$). 
Since there is an exponential decay in the Type-II models, it converges toward the $\Lambda$CDM result much faster than that in
the Type-I ones as the curvature increases.
Accordingly, it is possible to exist a much larger enhancement the Type-I model than
that in the Type-II one.

\begin{figure}[htbp]
\begin{center}
\includegraphics[width=0.49 \linewidth]{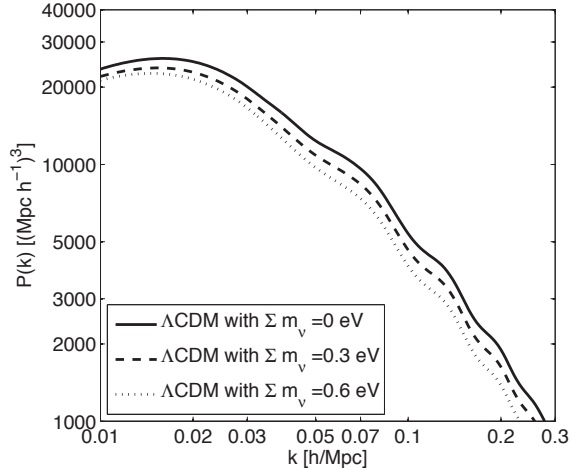}
\caption{
Matter power spectrum $P(k)$ in the $\Lambda$CDM model with $\Sigma m_{\nu}= 0$ (solid line),
0.3 (dashed line) and 0.6  (dotted line) $\mathrm{eV}$. }
\label{fg:3}
\end{center}
\end{figure}

To study the effect of massive neutrinos on the matter power spectrum, we first use the relation between the contributions
 of massive neutrinos to the total energy density, $\Omega_{\nu}$, and the total mass, $\Sigma m_{\nu}$, in unit of eV, given by
\begin{eqnarray}
\Omega_{\nu} \simeq \frac{\Sigma m_{\nu}}{94h^{2}\mathrm{eV}},
\end{eqnarray}
where $h$ is the reduced Hubble constant.
The upper value at $95\%$ C.L. for the
total neutrino mass constrained by the Planck data in the $\Lambda$CDM model~\cite{Ade:2013zuv}, $\Sigma m_{\nu} < 0.23 \ \mathrm{eV}$,
leads to $\Omega_{\nu} \leq 5 \times 10^{-3}$. Although it is a rather small ratio to the total energy density,
its effect on the matter power spectrum is still detectable, as mentioned in Sec.~1.
As demonstrated in Fig~\ref{fg:3},
we see that the free streaming massive neutrino suppresses the growth of $P(k)$ in the subhorizon scale~\cite{Lesgourgues:2006nd}.
 This effect of massive neutrinos on the matter power spectrum is opposite to that of $f(R)$ models~\cite{Motohashi:2010sj}.
 As a result,  the Type-I $f(R)$ models would allow a higher  scale for the total neutrino mass than Type-II.

\section{Constraints from Cosmological Observations}

We now perform the numerical simulations for the viable $f(R)$ models by
 using the CosmoMC package with some massive neutrinos and the latest cosmological data.
 The dataset is as follows: the CMB data from Planck
with both low-$l$ ($l < 50$) and high-$l$ ($l \ge 50$) parts
and WMAP  with only the low-$l$ one; the BAO data from BOSS DR11; the matter power spectral data
 from SDSS DR4 and WiggleZ Dark Energy Survey; and the SNIa data  from SNLS.
With this dataset, we explore the constraints on the Cold Dark Matter density, $\Omega_{c}h^{2}$,
and the sum of the active neutrino masses, $\Sigma m_{\nu}$, in the two viable $f(R)$ models.
In Table~\ref{table3}, we list the cosmological parameters  in our analysis.
\begin{table}[htbp]
\begin{center}
\caption{List of priors for parameters.
}
\begin{tabular}{|c||c|} \hline
Parameters & Priors
\\ \hline
Starobinsky model &   $1<\lambda< 7, n=2$
\\ \hline
Exponential model & $1<\beta<4$
\\ \hline
Baryon density & $5 \times 10^{-3}<\Omega_bh^2<0.1$
\\ \hline
CDM density & $10^{-3}<\Omega_ch^2<0.99$
\\ \hline
Neutrino mass & $0<\Sigma m_{\nu} < 1 $ eV
\\ \hline
Neutrino number (fixed)& $N_{\mathrm{eff}} =3.046$
\\ \hline
Neutrino number (varied)  & $2<N_{\mathrm{eff}}<6$
\\ \hline
Spectral index & $ 0.9 < n_s < 1.1$
\\ \hline
Reionization optical depth & $ 0.01 <\tau <0.8$
\\ \hline
\end{tabular}
\label{table3}
\end{center}
\end{table}

\begin{center}
\begin{figure}[htbp]
\includegraphics[width=0.49 \linewidth]{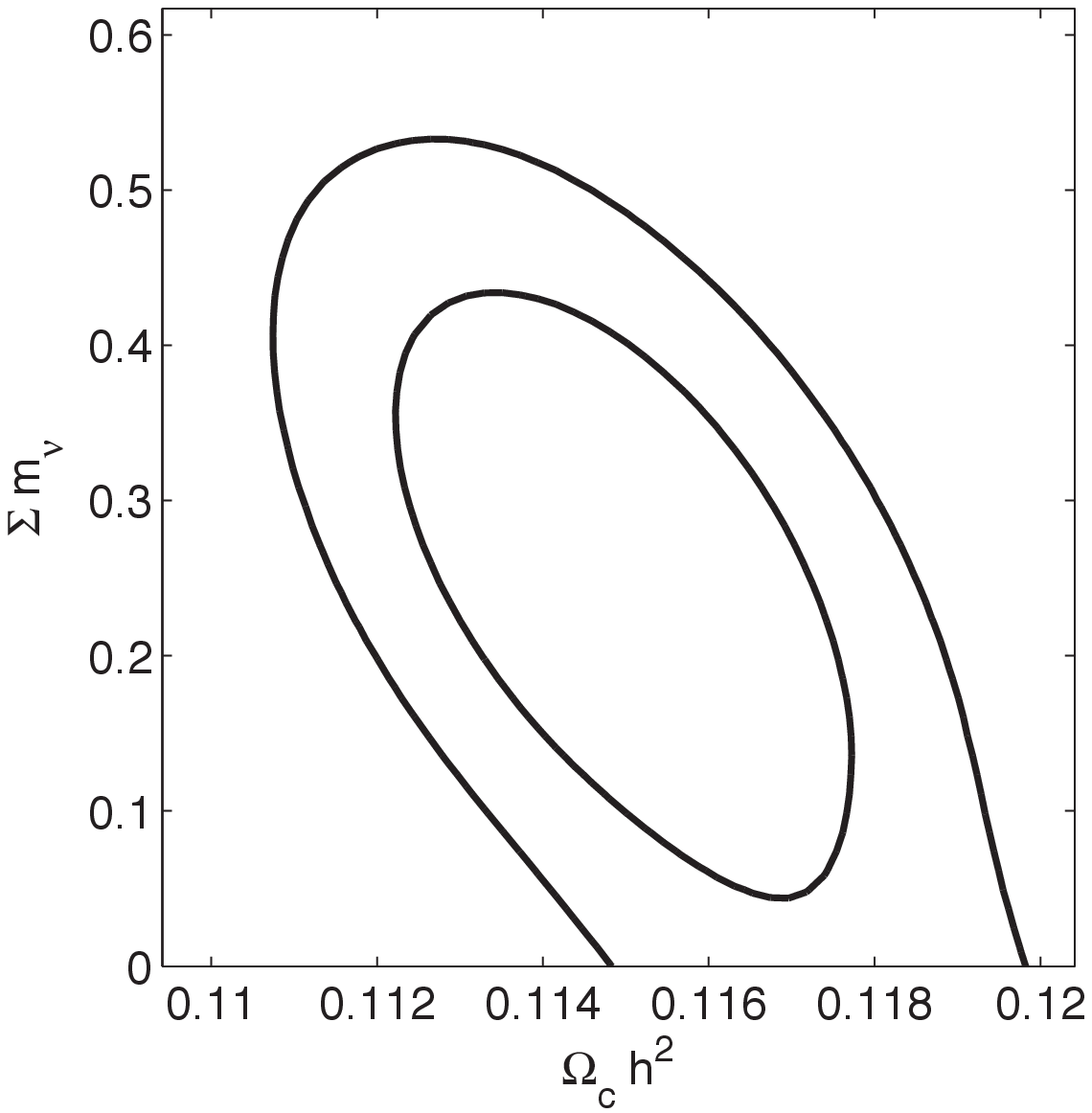}
\includegraphics[width=0.49 \linewidth]{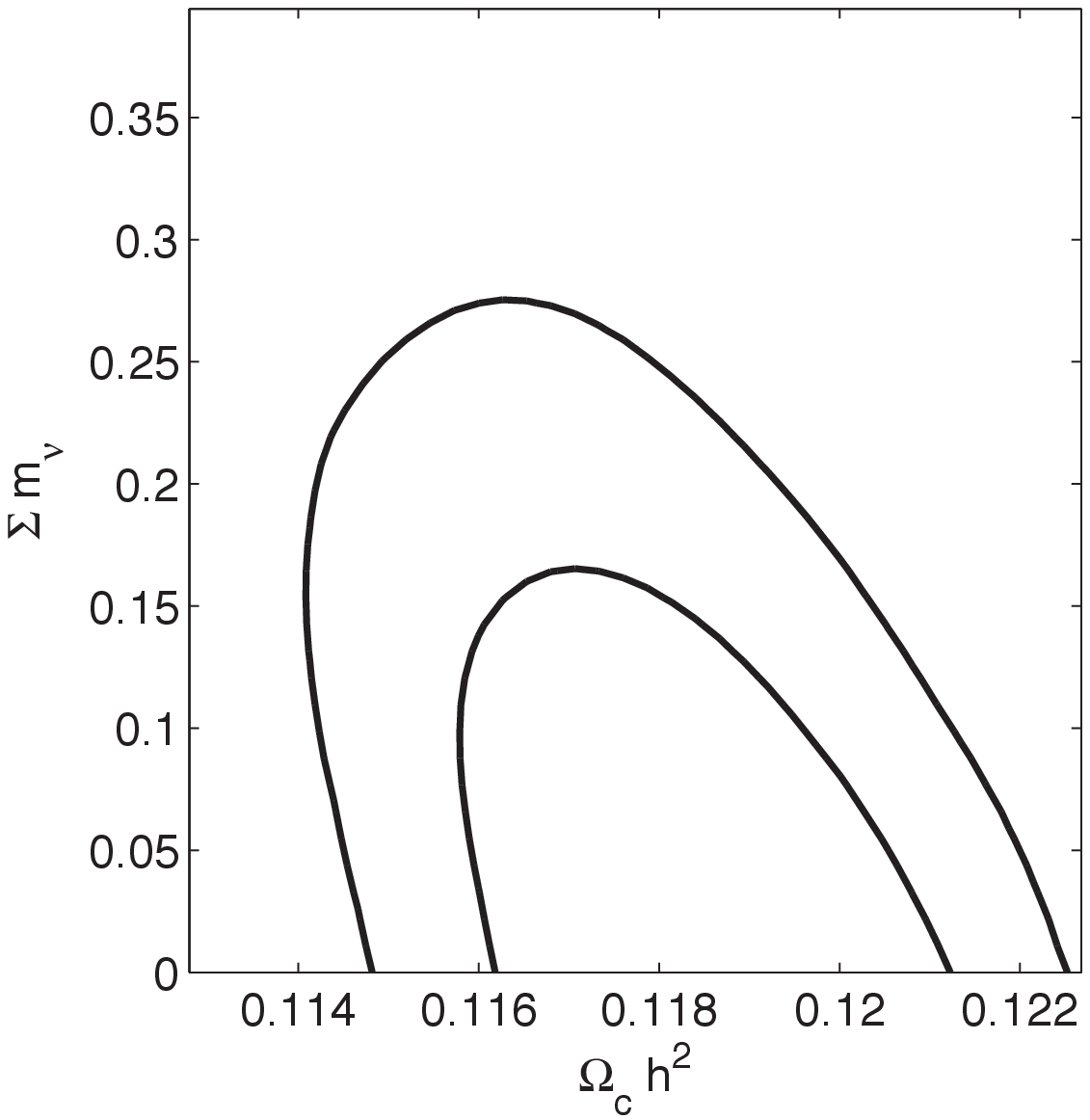}
\caption{
Contour plots of  $\Sigma m_{\nu}$ in $\mathrm{eV}$ and $\Omega_{c}h^{2}$
for the Starobinsky (left) and  exponential  (right) gravity models,
where the inner and outer curves represent $1\,\sigma$ and $2\,\sigma$ confidence levels, respectively.
}
\label{fg:5}
\end{figure}
\end{center}

In Fig.~\ref{fg:5}, we illustrate the constraining
contours of  $\Sigma m_{\nu}$ and $\Omega_{c}h^{2}$
with the assumption of only one massive neutrino.
Our results are summarized in Table~\ref{table1}.
\begin{table}[htbp]
\begin{center}
\caption{List of $\Sigma m_{\nu}$ and $\Omega_c h^2$ with $95\%$ C.L. in the $\Lambda$CDM,  Starobinsky
and exponential models.}
\begin{tabular}{|c||c|c|c|c|c|} \hline
$f(R)$ model & $\Sigma m_{\nu}$ & $\Omega_b h^2$ & $\Omega_c h^2$ & $n_s$ & $\tau$
\\ \hline
$\Lambda$CDM & $ < 0.200$ eV & $ 2.22^{+0.04}_{-0.06} \times 10^{-2}$ & $0.117^{+0.004}_{-0.002}$ & $ 0.963^{+0.010}_{-0.011} $ & $ 0.092 \pm 0.025 $
\\ \hline
Starobinsky & $ 0.248^{+0.203}_{-0.232}$ eV & $2.25^{+0.04}_{-0.05}\times 10^{-2}$ & $0.114^{+0.004}_{-0.002}$ & $0.971^{+0.09}_{-0.13}$ & $0.097^{+0.021}_{-0.030}$
\\ \hline
Exponential & $ < 0.214 $ eV & $ 2.22 \pm 0.05 \times 10^{-2}$ & $0.118 \pm 0.03$ & $0.964^{+0.009}_{-0.011}$ & $0.092^{+0.026}_{-0.025}$
\\ \hline
\end{tabular}
\label{table1}
\end{center}
\end{table}
 From Fig.~\ref{fg:5} and Table~\ref{table1}, we see that
 $\Sigma m_{\nu} < 0.451 \ \mathrm{eV}$ ($95\%$ C.L.) in the Starobinsky model,
 which is close to the upper bound given in Ref.~\cite{Dossett:2014oia},
 while the exponential model leaves the massive neutrino a rather small space, $\Sigma m_{\nu} < 0.214 \ \mathrm{eV}$ ($95\%$ C.L.),
 which is very close to that of the $\Lambda$CDM model, $\Sigma m_{\nu} < 0.200 \ \mathrm{eV}$ ($95\%$ C.L.).
 Clearly, the behavior  the Type-II viable $f(R)$ models  is barely distinguishable from the $\Lambda$CDM one.

In addition, we can  examine the effective number of neutrino species, $N_{\mathrm{eff}}$,
to account for the neutrino-like relativistic degrees of freedom,
defined by
\begin{eqnarray}
\rho_{\mathrm{radiation}} \equiv \rho_{\gamma} + \rho_{\nu} = \left[ 1 + \frac{7}{8}\left(\frac{4}{11}\right)^\frac{4}{3} N_{\mathrm{eff}} \right]\rho_{\gamma},
\end{eqnarray}
where $\rho_{\gamma}$ is the energy density of photon, $7/8$ comes from the Fermi-Dirac distribution since neutrinos
are fermions, and $4/11$ is due to the ratio of the neutrino to photon temperature.
The effect of $N_{\mathrm{eff}}$ is mostly on the epoch of the matter-radiation equality and the expansion rate as well as
the CMB power spectrum.

\begin{table}[htbp]
\begin{center}
\caption{List of $N_{\mathrm{eff}}$, $\Sigma m_{\nu}$ and $\Omega_c h^2$ with $95\%$ confidence levels in the $\Lambda$CDM,
 Starobinsky, and exponential models.}
\begin{tabular}{|c||c|c|c|c|c|c|} \hline
& $N_{\mathrm{eff}}$ & $\Sigma m_{\nu}$ & $\Omega_b h^2$ & $\Omega_c h^2$ & $n_s$ & $\tau$
\\ \hline
$\Lambda$CDM & $3.47^{+0.82}_{-0.47}$ & $ < 0.351$ eV & $2.24^{+0.06}_{-0.05} \times 10^{-2}$ & $0.125^{+0.012}_{-0.008}$ & $0.974^{+0.028}_{-0.015}$ & $0.086^{+0.038}_{-0.016}$
\\ \hline
Starobinsky & $3.78^{+0.64}_{-0.84}$ & $ 0.533^{+0.254}_{-0.411}$ eV & $2.28\pm 0.06 \times 10^{-2}$ & $0.124^{+0.010}_{-0.011}$ & $0.989^{+0.031}_{-0.026}$ & $0.099^{+0.030}_{-0.029}$
\\ \hline
Exponential & $3.47^{+0.74}_{-0.60}$ & $< 0.386$ eV & $2.26^{+0.04}_{-0.07} \times 10^{-2}$ & $0.124^{+0.011}_{-0.009}$ & $0.978^{+0.023}_{-0.022}$ & $0.092^{+0.032}_{-0.024}$
\\ \hline
\end{tabular}
\label{table2}
\end{center}
\end{table}

\begin{center}
\begin{figure}[htbp]
\includegraphics[width=0.49 \linewidth]{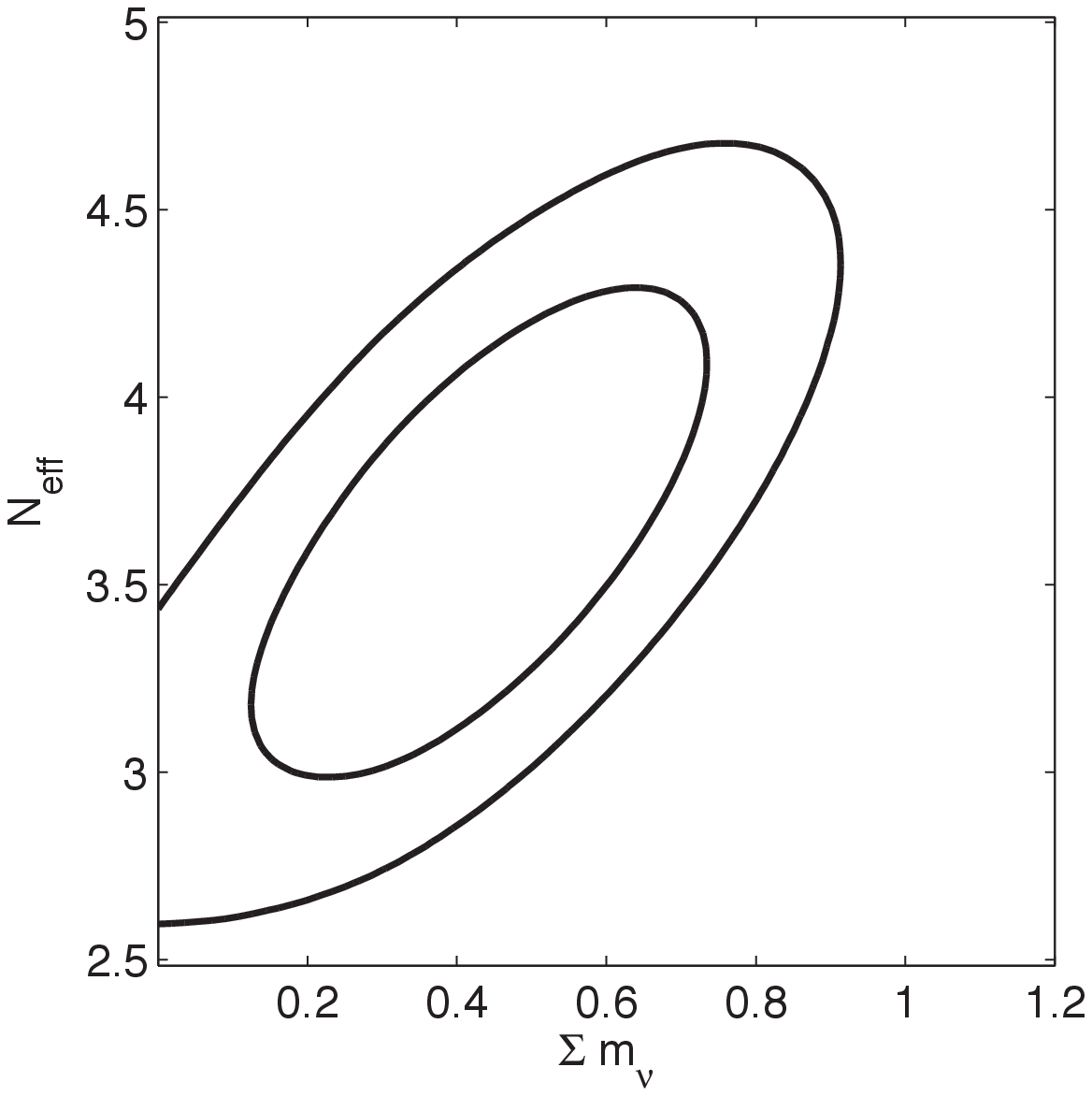}
\includegraphics[width=0.49 \linewidth]{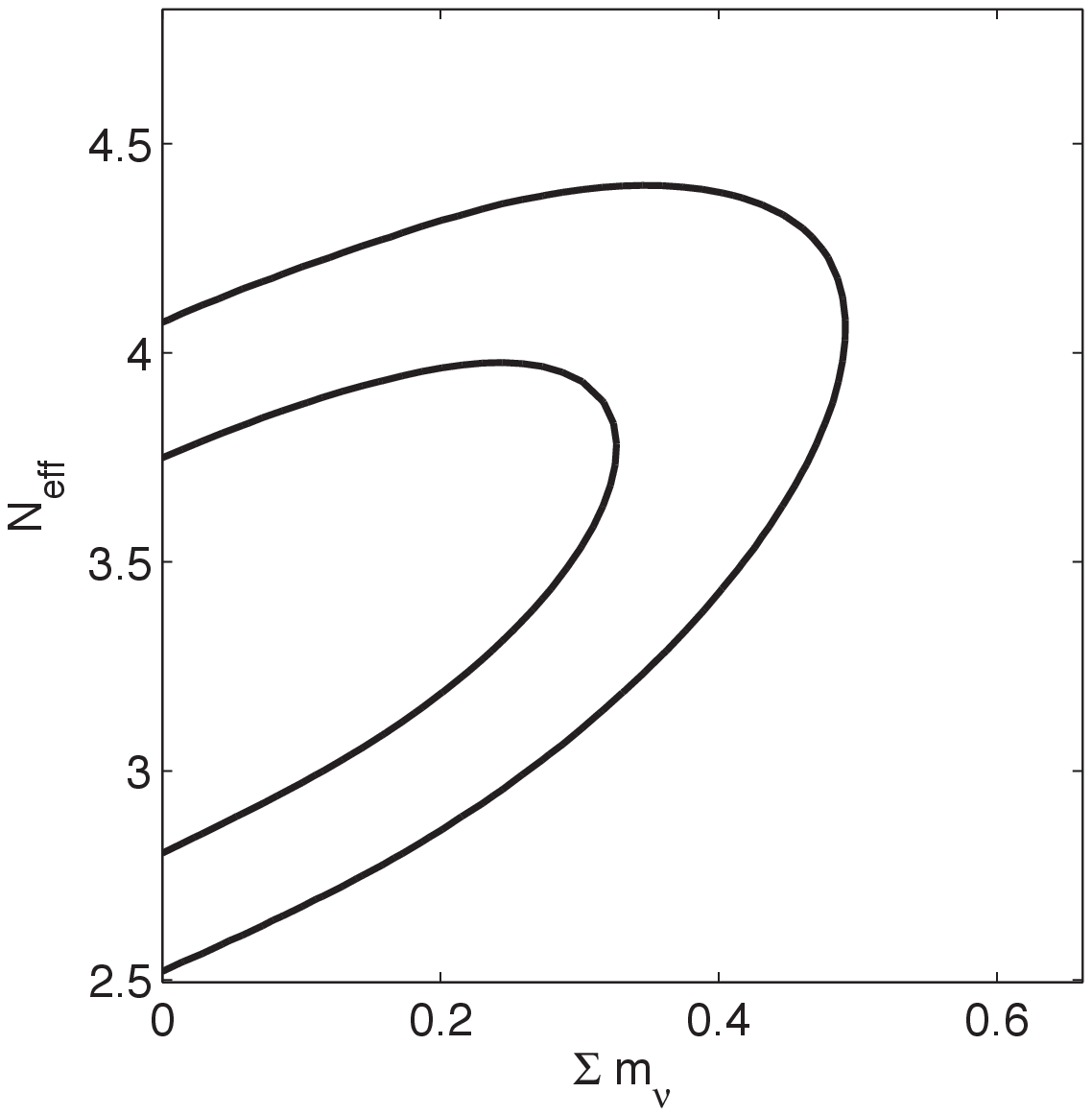}
\caption{Contours of  $N_{\mathrm{eff}}$ and $\Sigma m_{\nu}$ in the Starobinsky (left) and exponential (right) models
with $1\,\sigma$ and $2\,\sigma$ confidence levels, respectively.
}
\label{fg:6}
\end{figure}
\end{center}

Our results are  shown in Fig.~\ref{fg:6} and Table~\ref{table2}.
The best-fit value of $N_{\mathrm{eff}}$ is $3.78$
in the Starobinsky model,
which is higher than the corresponding one of $3.47$
 in both $\Lambda$CDM and exponential models.
 This infers that the Starobinsky model allows more relativistic species than the other two models.
 However, at $95\%$ C.L., the three models admit approximately the same range for $N_{\mathrm{eff}}$.
 On the other hand,  the total neutrino mass in the two viable $f(R)$ models as well and  the $\Lambda$CDM
 one  increases when $N_{\mathrm{eff}}$ is treated as a free parameter.

\section{Conclusions}

We have studied the effect of massive neutrinos in the two types of viable $f(R)$ gravity theories,
 the Starobinsky and  exponential models, by using the CosmoMC package with the modified MGCAMB.
We have considered the linear perturbations in these models by assuming that the background evolutions are
the same as the $\Lambda$CDM.
The enhancement of the matter power spectrum has been found to be a generic feature in the viable $f(R)$ gravity theory.
However, the biggest magnifications of the matter power spectra
 in the Type-I $f(R)$ models are more significant than those in the
Type-II ones.  With an increasing curvature, the results in the Type-I models approach to the $\Lambda$CDM as an inverse power law,
 while those in the Type-II models as an exponential decay.
 Clearly, the Type-I viable $f(R)$ models allow larger mass scales for massive neutrinos than those
in Type-II since massive neutrinos suppress the matter power spectra.
As a result,  the modified $f(R)$ gravity theory would be used to compensate the suppression from the effect of massive neutrinos.
If the three neutrinos are in a large mass scale, the Type-I viable $f(R)$ theory, such as
 the Starobinsky model, is favored.

Our investigation has shown that the allowed neutrino mass scale is further released when $N_{\mathrm{eff}}$
 is considered as a free parameter. Moreover, the best-fit value of $N_{\mathrm{eff}}$ has been found to be $3.78$ in the Starobinsky model,
 which is greater than 3.47 in the $\Lambda$CDM and exponential models.
  Clearly, the Starobinsky model leaves  more room for a dark sector or a sterile neutrino in the universe.

\begin{acknowledgments}
The work was supported in part by National Center for Theoretical Sciences, National Science
Council (NSC-101-2112-M-007-006-MY3) and National Tsing Hua
University (103N2724E1).
\end{acknowledgments}

\end{document}